\documentclass{icrc2009}

\usepackage{graphicx}   
\usepackage{url}

\newcommand{\shorttitle}[1]%
{\markboth{Proceedings of the 31\MakeLowercase{$^{st}$} ICRC, {\L}\'{o}d\'{z} 2009}{#1} }
\newcommand{\etal}{\MakeLowercase{\textit{et al. }}} 

\begin{document}
\title{On the possible connection between cosmic rays and clouds}

\author{\IEEEauthorblockN{Anatoly Erlykin\IEEEauthorrefmark{1},
			  Gyula Gyalai\IEEEauthorrefmark{2},
                          Karel Kudela\IEEEauthorrefmark{2},
                           Terry Sloan\IEEEauthorrefmark{3} and
                           Arnold Wolfendale\IEEEauthorrefmark{4}}
                            \\
\IEEEauthorblockA{\IEEEauthorrefmark{1}P.N.Lebedev Physical Institute, Moscow, Russia}
\IEEEauthorblockA{\IEEEauthorrefmark{2}Institute of Experimental Physics, Kosice, 
Slovakia}
\IEEEauthorblockA{\IEEEauthorrefmark{3}Department of Physics, Lancaster University, 
Lancaster, UK}
\IEEEauthorblockA{\IEEEauthorrefmark{4}Department of Physica, Durham University, 
Durham, UK}}

\shorttitle{Erlykin \etal cosmic rays and clouds}
\maketitle

\begin{abstract}
Various aspects of the connection between cloud cover (CC) and cosmic rays (CR) are
analysed. We argue that the anticorrelation between the temporal behaviour of low (LCC)
 and middle (MCC) clouds evidences against the causal connection between them and CR. 
Nevertheless, if a part of low clouds (LCC) is connected and varies with CR, then its 
most likely value averaged over the Globe should not exceed 20\% at the two standard 
deviation level. 
\end{abstract}

\begin{IEEEkeywords}
correlation, cosmic rays, clouds
\end{IEEEkeywords}
 
\section{Introduction}
A correlation between CR intensity and global LCC was observed for the first time more 
than 10 years ago \cite{Sven1,Palle} and led to a new direction in 
science - cosmoclimatology \cite{Sven2}. It is based on the concept of a causal 
relationship 
between CR and CC. Some arguments against this causality were presented in \cite{Sloan}
. The purpose of this study is to continue further an analysis of possible reasons for 
the observed correlation between CR and LCC.
\section{Input data}
 As the input CC data, we took the same observations with weather satellites (~ISCCP 
project~) that were used in \cite{Sven1,Palle}. We analysed sky fractions covered by 
clouds, averaged over observation months (D2 series). In compliance with the ISCCP 
cloud height classification made according to the pressure at their upper boundary, 
clouds were separated into low (LCC,$>$680 hPa), medium (MCC, 440-680 hPa) and high 
(HCC,$<$440hPa). 
Because of the ongoing discussion on the calibration quality of ISCCP radiometers
after 1996 \cite{Marsh} we used both the data obtained earlier, during the 22nd cycle 
of solar activity (July 1986 - December 1995), and the complete set of data, till 2005.
 For comparison of CC and CR variations we used the data from several neutron monitors 
of the worldwide network (Thule, Apatity, Moscow, Climax, Huancayo). \\
 In an analysis of the latitude dependence of CR and LCC variations we split the entire
 latitude range (from $-90^\circ$ to $90^\circ$) into nine equal intervals of 
$20^\circ$ width. We also analysed the dynamics of global (i.e. globe-averaged) CC. 
To better
reveal the CC variations of nontrivial origin, we subtracted the winter-summer 
seasonal variations from the temporal curves, although in some cases seasonal 
variations were considered as well.
\section{Results}
\subsection{Long-term variations and the fraction of LCC correlated with CR intensity}
\begin{figure}[hptb]
\centering
\includegraphics[width=3.4in]{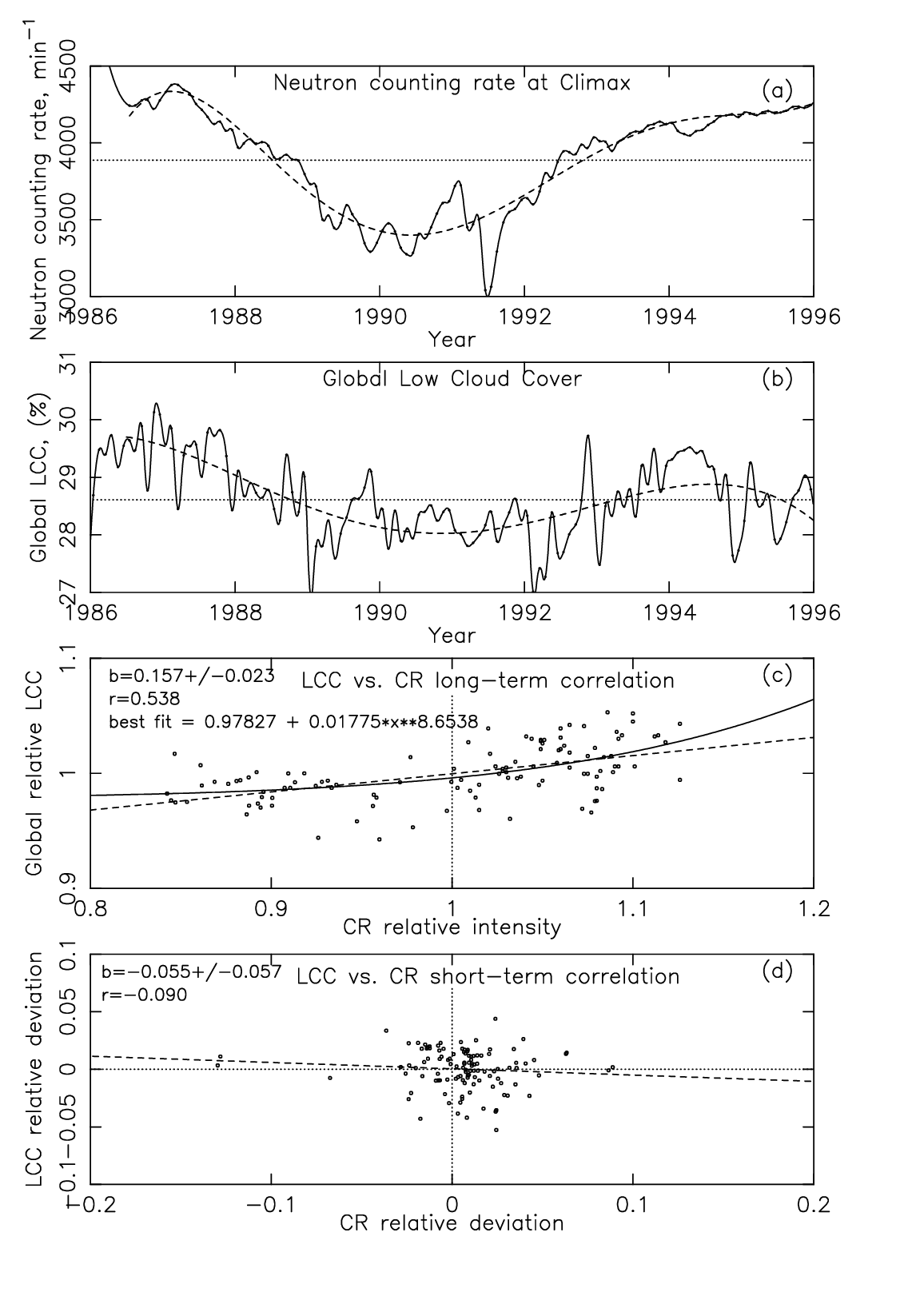}
\caption{\footnotesize CR and LCC variations and their correlation during the 22nd 
cycle of solar activity. 
(a) CR variation (counting rate in the Climax neutron monitor); (b) global LCC
 variation; (c) correlation of CR and LCC variations with respect to their average 
levels over the period 1986-1996 (long-term correlation), $b = 0.157 \pm 0.023,
 r=0.538$, the best fit is $0.978 + 0.018x^{8.65}$; (d) correlation of CR and LCC 
variations with respect to the average temporal evolution curve - thin smooth lines in 
Figures 1a and 1b (short-term correlation), $b = -0.055 \pm 0.057, r = -0.0904$. The 
short-dash lines in all panels are the average levels over the period 1986-1996, the 
long-dash lines in Figures 1c and 1d are linear regression lines with a slope $b$, 
the thin smooth line in Fig.1c is the best power-law fit to the scatter plot, and $r$ 
is the correlation coefficient.}
\label{Fig01}
\end{figure}
Figure 1 shows the temporal evolution of (a) CR intensity, (b) global LCC and (c) the 
correlation of CR and LCC deviations from their means. To illustrate the CR intensity 
evolution we took as a proxy the data of the Climax monitor, which is situated at a 
latitude of $39.4^\circ N$. The CR intensity fluctuations at other latitudes show 
qualitatively similar temporal behavior, although with different variation amplitudes. 
The CR and LCC deviations from their means are positively correlated. The linear 
regression slope is 0.157$\pm$0.023 and the corresponding correlation coefficient is 
0.538, which confirms the existence of positive correlation between CR and LCC that was
 found in \cite{Sven1,Palle}. \\
    However, an attempt to understand what is the reason and what is the consequence 
here failed. We wanted to find a possible time shift between the CR and LCC 
variations by the method of least squares. However, it turned out that the sum of 
squared deviations has a flat broad minimum at a respective shift of the CR and LCC 
curves from -11 to +6 months. That is , one cannot say which variation is the cause and
 which is the consequence. \\
    On the assumption that CR are indeed responsible for at least some part of the CC 
one can use the observed correlation to estimate this part. This estimation depends on 
the model for the relationship between CR and LCC. With a linear model 
($\Delta / \langle \Delta \rangle = a + b(I/ \langle I \rangle)^c$, 
where $c = 1$; $\Delta$ and $I$ are the cloud coverage and CR intensity respectively;
and $\langle \Delta \rangle$ and $\langle I \rangle$ are their mean values ), 
the regression slope $b = 0.157$ gives the CC fraction related to CR, to be 
approximately 16\%. Within two standard deviations this fraction should not exceed 
20\%. However, a least-squares estimation of the parameters $a, b$ and $c$ shows that 
the relation between CR and LCC is most likely non-linear. We obtained the values 
$a = 0.978, b = 0.018$ and $c = 8.65$ ( thin solid line in Fig.1c ). This shows that 
the most likely CC fraction related to CR does not exceed 2\%. \\
    The above conclusion is valid only if the models of relationship between CR and 
LCC are correct and the CR variations at the Climax latitude, $-40^\circ N$, adequately
 describe the global variation picture. For $c<1$ the CC fraction which positively 
correlates
with CR can be much higher. Unfortunately, because of the relatively small magnitude of
CR and LCC variations one cannot reliably estimate the parameter $c$ and thus evaluate 
more precisely the positively correlated CC fraction. Although the method of least 
squares is more adequate at $c>1$, in the domain of available experimental data 
(Fig.1c) the behavior of the curves for different values of $c$ is similar, as well as 
the corresponding sums of the squared deviations.
\subsection{Short-term variations}
Figure 1 and the analysis given in the previous section concern the total variations of
 CR and LCC relative to their average values. The main contribution to these variations
 is from the long-term variations related to the 11-year solar activity cycle. To 
reveal 
possible correlations in short-time CR and LCC variations, we excluded the contribution
 from long-term variations. For this purpose we approximated the temporal evolution of 
CR and LCC by a fifth-order polynomial (smooth solid lines in Figures 1a and 1b) and 
calculated deviations from this approximation. Because we used the monthly averaged 
D2-series data, this analysis is related to monthly variations. We did not find any 
statistically significant correlations between the CR intensity and global LCC. The 
estimated regression line slope is $b=-0.055\pm 0.057$ and the corresponding 
correlation 
coefficient is -0.090 (Fig.1d). This negative result is in indirect agreement with the 
absence of even shorter (few days) correlations ( Forbush decreases, CR ground-level
enhancements ) pointed out in \cite{Sloan, Krist}.
\subsection{Latitude dependence of correlations between CC and CR}
Because the CR intensity depends on the latitude it is reasonable to analyse the 
variation of CC characteristics with latitude. 
\begin{figure}[hptb]
\centering
\includegraphics[width=3.5in,height=3in]{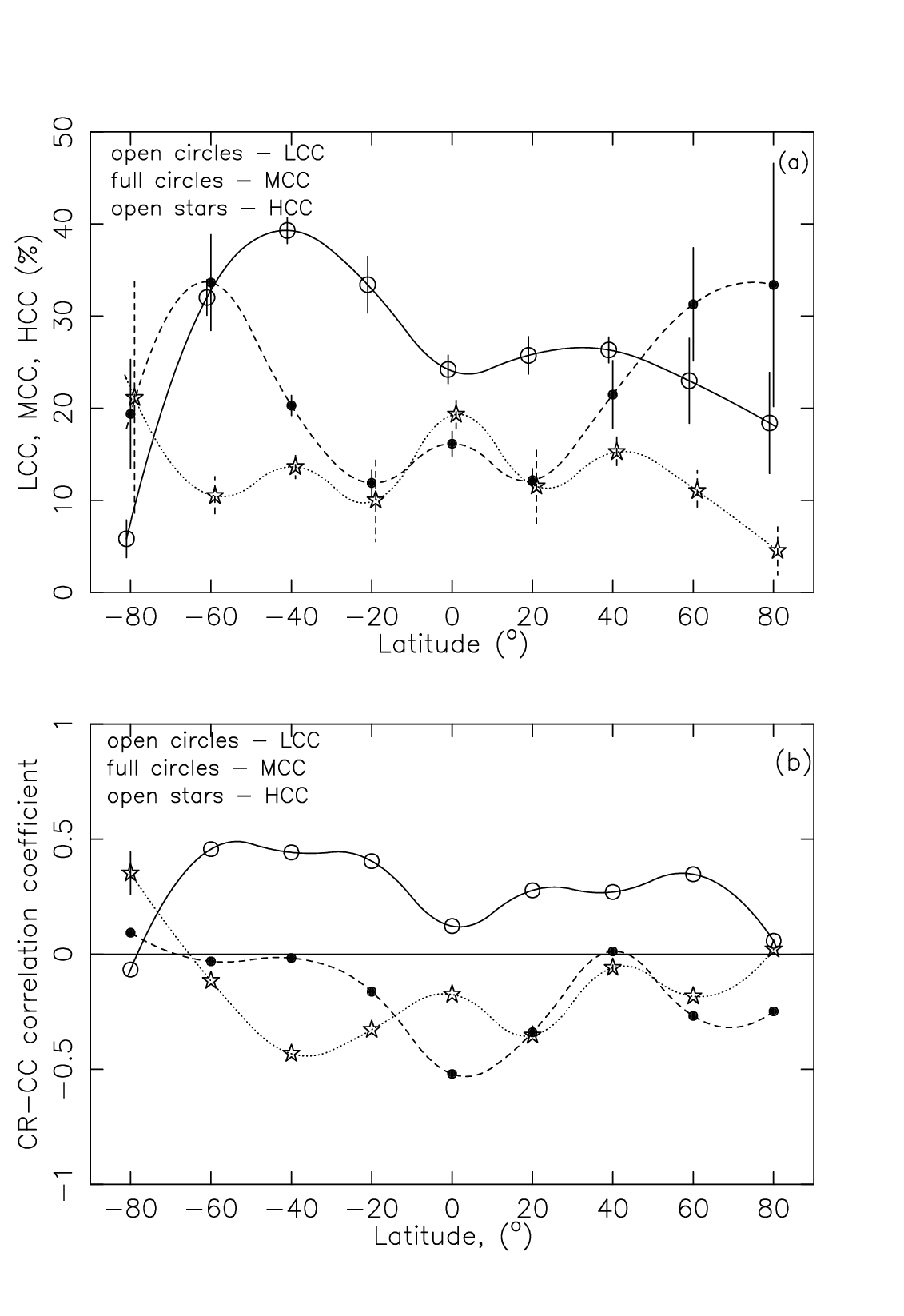}
\caption{\footnotesize The latitude dependence of CC characteristics: (a) absolute 
values of LCC 
(open circles), MCC (full circles) and HCC (open stars). (b) LCC, MCC and HCC 
correlations with CR (Climax). Notations are the same as in (a).}
\label{Fig02}
\end{figure}
Figure 2a shows the latitude dependence of LCC, MCC and HCC. It is seen that there is a
 small minimum for LCC in the equatorial region, which could be connected with the 
reduction of CR intensity, but it is not confirmed by the local maxima in MCC and HCC.
In polar regions, where the CR intensity is the highest, there is an opposite decrease 
of LCC, which apparently is connected with the dominant influence of the atmospheric 
conditions, eg. low temperatures. The highest LCC is in the southern latitude bands 
with the largest part of the area occupied by oceans, i.e. with a relatively large 
density of water vapor. \\
The altitude dependence of CC does not correspond to the altitude dependence of the CR 
intensity: a further bad feature. In most of the latitude bands, MCC and HCC are 
smaller than LCC, which is 
opposite to CR with their intensity rising with height. All this shows that even if
 there is a causal connection between CR and LCC, its character is more complicated 
than a direct and positive correlation. \\
We have already mentioned that the global LCC-CR correlation is positive: $r = 0.538$.
Figure 2b shows the latitude dependence of the CC-CR correlation coefficient. Here 
again we used as a proxy of the CR temporal variations just the neutron counting rate 
at Climax. In spite of the latitude 
dependence of the CR variation amplitude, the value of the LCC-CR correlation 
coefficient does not depend on this amplitude due to the similarity of the temporal 
behavior of CR variations at different latitude bands. It is remarkable to note that in
 most latitude bands, MCC and HCC have negative correlation with CR in contrast to the
positive LCC-CR correlation which was the main argument for the claimed causal CR-CC 
connection \cite{Sven1,Palle}. 
\subsection{Negative correlations of LCC and MCC}
Figure 3 shows the latitude dependence of the sensitivity and correlation between MCC 
and LCC.
\begin{figure}[hptb]
\centering
\includegraphics[width=1.8in,height=3.5in,angle=-90]{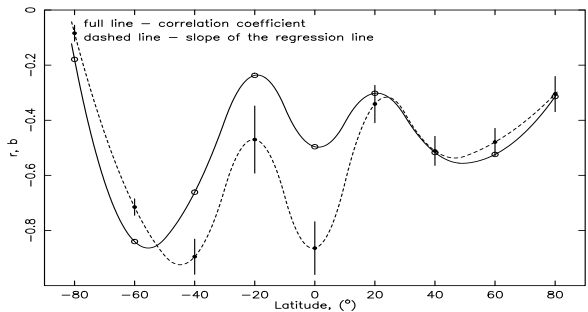}
\caption{\footnotesize The latitude dependence of the sensitivity (slope $b$ of the 
linear regression line) of MCC to LCC variations (full line) and their correlation 
coefficient $r$ (dashed line).}
\label{Fig03}
\end{figure}
 The sensitivity of one variable to another, according to the definition 
\cite{Uchai}, is the derivative of the first variable on the second in log-coordinates.
 In our case the sensitivity is the slope of the linear regression line $b$ in the 
MCC-LCC plot. One can notice two features: \\ 
(i) the sensitivity of MCC to LCC and MCC-LCC correlation coefficient are 
negative at all latitudes, which is another support of their global 
anticorrelation. The negative sensitivity of MCC to LCC is difficult to explain in the 
framework of the causal connection between CC and CR, since the rise of the CR 
intensity has to change CC similarly at all altitudes; \\
(ii) the highest negative sensitivity and the correlation between MCC and LCC is 
observed in tropical and subtropical regions $\ell = -30^\circ / +30^\circ$ as well as 
in southern latitude bands with the highest fraction of water: 
$\ell = -45^\circ / -65^\circ$.
\section{Discussion}
In our opinion the analysis performed here, as well as our previous arguments 
\cite{Sloan}, gives grounds to assert that CR are not the dominant factor leading to CC
 formation. The negative correlation of LCC and MCC, most prominent in the tropics and 
subtropics and at the latitudes where one can expect excess water vapor formation, 
allows one to turn to the traditional picture describing the main reasons for cloud
 formation that are connected with solar activity. \\
\begin{figure}[hptb]
\centering
\includegraphics[width=3.4in]{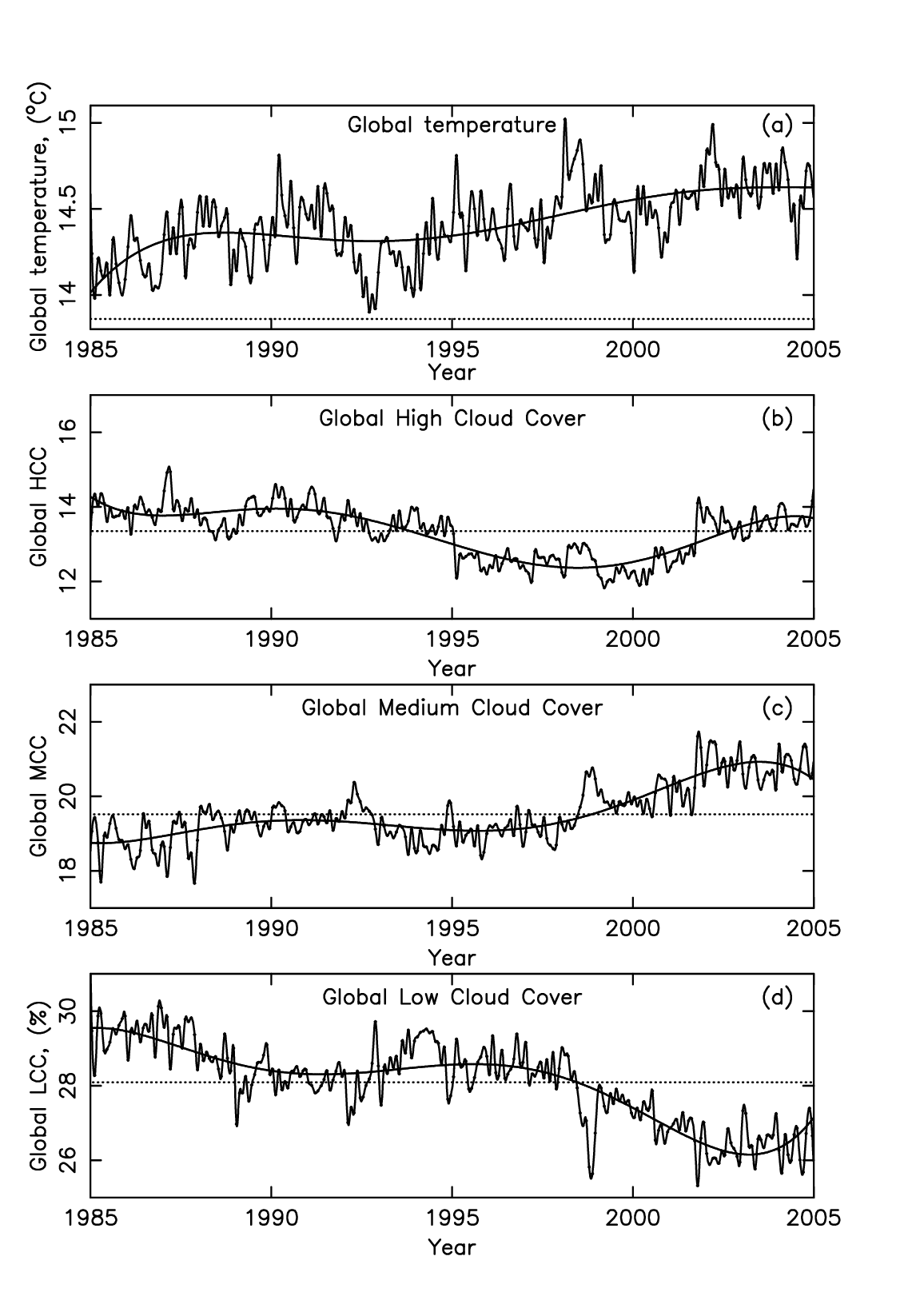}
\caption{\footnotesize Temporal evolution of the surface temperature (a), HCC (b), MCC (c) and LCC 
(d) during two last cycles of solar activity. The dotted line in panel (a) is the 
average temperature in the 20th century. Dotted lines in panels (b)-(d) show the 
average CC level over the measurement period 1984-2005.}
\label{Fig04}
\end{figure}
Solar radiation increases together with the number of sunspots in the middle of the 
solar cycle. This radiation is strongest in the tropics and subtropics. Although the 
relative increase in radiation intensity is insignificant ($\sim$0.1\%), it leads to an
 increase in the average ground-level temperature and enhances vertical convective 
flows of heated air. Since the cloud height in the ISCCP experiment is classified 
according to the pressure at the upper cloud boundary, the convective lift of clouds to
 higher altitudes leads to a redistribution of the assigned altitudes. That is, it 
decreases LCC and increases MCC, which is reflected in the negative correlations of LCC
 and MCC. Thus, an increase in convective flows leads to a significant strengthening 
($\sim$2\%) of the effect of increased solar radiation. \\
Along with the periodic variations of the ground-level Earth's temperature, related to 
the periodicity of solar cycles, a systematic temperature increase due to global 
warming was observed in the last century. The warming was particularly strong during 
the last two decades. Figure 4a shows the average temperature growth, and Figures 4b-4d
demonstrate the temporal evolution of HCC, MCC  and LCC, respectively. 
 One can see both
 periodic and systematic variations of MCC and LCC, which are negatively correlated. 
The data on the graphs suggest that the tendency of LCC reduction, which appeared 
after 1995 \cite{Marsh} is not due to malfunctioning of the weather satellite equipment
 but due to a global warming of the Earth's climate. \\
Summing up, one can state that with an increase in the ground-level Earth's temperature
 during solar activity maxima or general global warming, the average CC height becomes
somewhat greater. This effect is most prominent in the tropics, subtropics and above the 
\newpage
surface of the ocean. The simultaneous reduction of LCC and of CR intensity is not 
evidence for a causal relationship between these two phenomena. They correlate due to 
the presence of a common driving force: changes in solar activity. \\
{\em Acknowledgments} \\
K.Kudela wishes to acknowledge the VEGA grant agency, project 2/7063/27. A.D.Erlykin 
acknowledges the support of the John C. Taylor Charitable Foundation and is grateful to
 E.A.Vashenyuk for granting access to the Apatity neutron monitor data and to S.P.Perov
 for useful discussions.

\end{document}